\documentstyle[epsfig]{aipproc}

\begin{document}
\title{The Trans-Relativistic Blast Wave Model\\ for SN 1998bw and GRB 980425}
\author{Jonathan C. Tan$^1$, Christopher D. Matzner$^2$,\\ and Christopher F. McKee$^{1,3}$}
\address{$^1$Dept. of Astronomy, University of California, Berkeley, CA 94720, USA\\
$^2$CITA, University of Toronto, 60 St. George Street, Toronto, Ontario M5S 3H8, Canada\\$^3$Dept. of Physics, University of California, Berkeley, CA 94720, USA}

%\lefthead{LEFT head}
%\righthead{RIGHT head}
\maketitle

\begin{abstract}
The spatiotemporal coincidence of supernova (SN) 1998bw and gamma-ray
burst (GRB) 980425 and this supernova's unusual optical and radio
properties have prompted many theoretical models that produce GRBs
from supernovae. We review the salient features of our simple,
spherical model in which an energetic supernova explosion shock
accelerates a small fraction of the progenitor's stellar envelope to
mildly relativistic velocities. This material carries sufficient
energy to produce a weak GRB and a bright radio supernova through
an external shock against a dense stellar wind.
\end{abstract}

Crack! A whip's tail flies supersonically through the air after a slow
transverse wave has accelerated down its tapering length. In the same
manner a shock wave propagating through a star may accelerate to
relativistic speeds in the tenuous layers of the stellar atmosphere
\cite{col74,mat99,tan01}. Furthermore, a pressure gradient, imprinted
in the expanding postshock gas, greatly boosts the kinetic energy of
the outer ejecta. Large energies are naturally concentrated in small
amounts of matter since shock acceleration only occurs at steeply
declining density gradients. Supernova-driven shock acceleration
thus provides an attractive model for gamma-ray bursts (GRBs)
\cite{col74}, one that naturally overcomes the baryon-loading problem.

To quantify this process, we have performed a suite of numerical
simulations of the acceleration of blast waves from nonrelativistic to
ultrarelativistic velocities in simple, idealized density
distributions \cite{tan01}. One of our principal results is an
analytic expression for $E_k(>\Gamma\beta)$, the kinetic energy
of ejecta moving with final velocity greater than $\beta\equiv v/c$ and
Lorentz factor greater than $\Gamma\equiv(1-\beta^2)^{-0.5}$, from
explosions in centrally concentrated density distributions with
polytropic envelopes. We applied this analysis to a more realistic,
though still one dimensional, supernova progenitor, kindly provided by
Stan Woosley, that was matched to SN 1998bw's light
curve\cite{woo99}. It involves a $2.8\times 10^{52}\:{\rm erg}$
explosion in a Wolf-Rayet (WR) star consisting of the $\sim 7\:{\rm
M_\odot}$ carbon-oxygen core of an initially $\sim 25\:{\rm M_\odot}$
main sequence star. We also simulated this explosion with our
relativistic code. Both our analytic and numerical estimates place
$\sim 2\times 10^{48}\:{\rm ergs}$ in ejecta with $\Gamma\beta>1$,
sufficient to account for GRB 980425, which has isotropic
$E_\gamma\simeq 8.5\times 10^{47}\:{\rm ergs}$.

We find that $E_k(>\Gamma\beta)\propto E_{\rm in}^{2.67\gamma_p}M_{\rm
ej}^{1-2.67\gamma_p}$, where $E_{\rm in}$ and $M_{\rm ej}$ are the
total kinetic energy and mass of the ejecta and $\gamma_p\equiv
1+1/n$, with $n$ being the polytropic index of the outer stellar
envelope that makes up the relativistic ejecta. SN~1998bw, classified
as a peculiar Type Ic and thus lacking H and He, is an energetic
explosion with a relatively small total ejecta mass. This explains its
ability to produce relativistic ejecta. However, whether a given supernova
creates a detectable GRB or not also depends on the properties of the
circumstellar wind (see below).

How do the more energetic cosmic GRBs relate to this model for
GRB~980425? $E_k(>\Gamma\beta)$ depends sensitively on $E_{\rm in}$,
which itself depends on the gravitational energy release,
$E_{\rm grav}$, and the efficiency of coupling to the ejecta,
$\epsilon_{\rm in}\equiv E_{\rm in}/E_{\rm grav}$. Our model for
SN~1998bw forms a $1.8\:{\rm M_\odot}$ neutron star, but 
much more energetic explosions (``hypernovae'') may result
from the formation of some, more massive black holes \cite{pac98}. 
While estimates of $\epsilon_{\rm in}$ are
uncertain, explosions ejecting orders of magnitude more kinetic
energy than SN~1998bw are astrophysically conceivable and need occur
only rarely to account for the observed cosmic GRBs. Of course the limit
$E_k\leq E_{\rm in}$ truncates the above scaling. Our
preliminary analysis of cosmic GRBs in the
context of shock acceleration models, suggests that asymmetric
explosions, perhaps resulting from rotationally flattened progenitors,
are required \cite{tan01}.

\begin{figure}[t!]
\centerline{
\epsfig{file=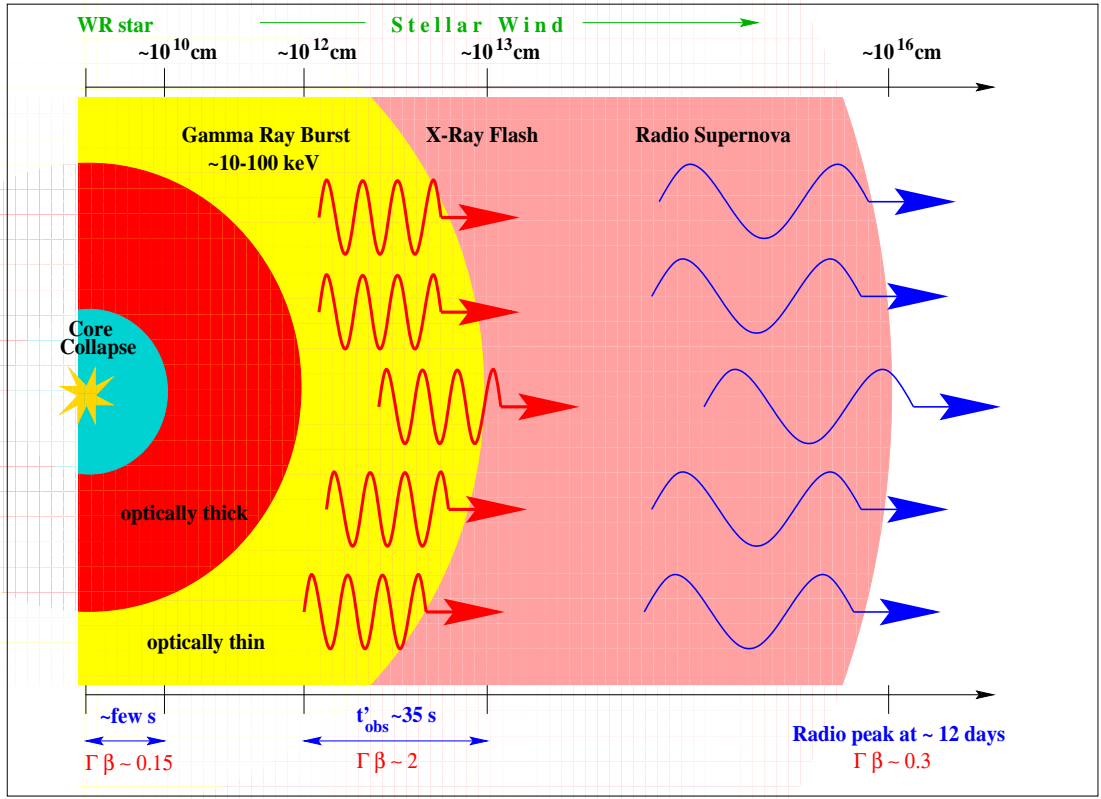,height=6in,width=4.5in,angle=-90}
}
\caption{SN 1998bw - GRB 980425 schematic.}
\label{fig1}
\end{figure}

Fig. \ref{fig1} summarizes our model of SN 1998bw and GRB 980425. The
energy of the mildly relativistic ejecta is greater than the observed
$\gamma$-rays, but to liberate this energy as radiation the ejecta
must collide with circumstellar material; in this case a stellar
wind. The $\sim 35\:{\rm s}$ timescale of the GRB sets the required
density of the wind since relativistic ejecta lose $\sim1/2$ of their
energy after sweeping up $1/\Gamma$ of their own mass. This wind density
is high: $\dot{M}_{-4}/v_{w,8}\sim 3$, where the mass loss rate is
$\dot{M}_{-4}=\dot{M}/10^{-4}\:{\rm M_\odot\:yr^{-1}}$
and the wind velocity is $v_{w,8}=v_w/10^8\:{\rm cm\:s^{-1}}$. In
fact some carbon rich Wolf-Rayet stars do lose mass at these extreme rates
\cite{koe95}.

For a given shock $\Gamma$ and postshock magnetic energy fraction, the
energy of synchrotron photons emitted by electrons behind the shock
front increases with density. For the mildly relativistic ejecta
produced in our model of SN 1998bw, the high densities implied by the
short GRB timescale are also required to give the correct energies
($\sim 10-100\:{\rm keV}$) of the GRB photons. Note, however, that
inverse Compton scattering may also affect the burst
spectrum. Interestingly, the high wind densities imply that at
the start of the interaction producing the burst, the wind is
optically thick to the $\gamma$-rays, while at the end it is optically thin. The energy
dependence of the Klein-Nishina cross section then predicts the early
emergence of the hardest photons from the wind, as was observed
\cite{pia00}.

Radio scintillation observations of the supernova at 12 days imply a
mean shock expansion velocity $\gtrsim 0.3 c$ over this period
\cite{kul98}. An external shock driven by ejecta from our model
explosion satisfies these constraints. The shock decelerates only very
slowly as the interior ejecta contain progressively more mass, while
the wind becomes ever more tenuous. At $\sim 20$ days the radio light
curve brightens, and this has been modeled as resulting from additional
energy injection into the shock front \cite{li99}. 
A natural explanation is that inner mass shells of ejecta, with
smaller energy per baryon but greater total energy than the outer layers,
eventually catch up to the
decelerating shock front. The relatively abrupt brightening
may result from a density shelf initially present in the
star's atmosphere, $\sim 10^{-3}\:{\rm M_\odot}$ from the
surface.

\end{document}